\documentstyle[11pt,epsf,aaspp4]{article}
\lefthead{F. Frontera et~al.}
\righthead{Spectral properties of GRB970228}

\def\S{{\em BeppoSAX\/}}   \def\eg{{\em e.g.,\ }}
\def\ergcm{\mbox{ erg cm$^{-2}$}}
\def\ergcms{\mbox{ erg cm$^{-2}$ s$^{-1}$}}

\makeatletter
\def\@cite#1#2{(#1\if@tempswa , #2\fi)}
\def\preprint{preprint}   \newif\ifPreprintMode
\ifx\preprint\revtex@genre\PreprintModetrue\else\PreprintModefalse\fi
\makeatother

\begin{document}

\title{Spectral Properties of the Prompt X--ray emission
 and Afterglow from the Gamma--Ray Burst of 28 February 1997}

\author{F. Frontera\altaffilmark{1,2}
E.~Costa\altaffilmark{3},
L.~Piro\altaffilmark{3},
J.M.~Muller\altaffilmark{4,5},
L.~Amati\altaffilmark{3},
M.~Feroci\altaffilmark{3},
F.~Fiore\altaffilmark{5},
G.~Pizzichini\altaffilmark{1},
M.~Tavani\altaffilmark{6,7},
A.~Castro-Tirado\altaffilmark{8},
G.~Cusumano\altaffilmark{9},
D.~Dal~Fiume\altaffilmark{1},
J.~Heise\altaffilmark{4},
K.~Hurley\altaffilmark{10},
L.~Nicastro\altaffilmark{1},
M.~Orlandini\altaffilmark{1},
A.~Owens\altaffilmark{11},
E.~Palazzi\altaffilmark{1},
A.N.~Parmar\altaffilmark{11},
J.~in~'t~Zand\altaffilmark{4},
G.~Zavattini\altaffilmark{2}
}

\altaffilmark{1}{Istituto Tecnologie e Studio Radiazioni Extraterrestri, 
CNR, Via Gobetti 101, 40129 Bologna, Italy}

\altaffilmark{2}{Dipartimento di Fisica, Universit\`a di Ferrara, Via Paradiso
 12, 44100 Ferrara, Italy}

\altaffilmark{3}{Istituto Astrofisica Spaziale, C.N.R., Via E. Fermi 21,
  00044 Frascati, Italy}

\altaffilmark{4}{Space Research Organization in the Netherlands,
 Sorbonnelaan 2, 3584 CA Utrecht, The Netherlands}

\altaffilmark{5}{\S\ Scientific Data Center, Via Corcolle 19, 00131 Roma,
 Italy} 

\altaffilmark{6}{Istituto Fisica Cosmica e Tecnologie Relative, C.N.R.,
  Via Bassini 15, 20133 Milano, Italy} 

\altaffilmark{7}{Columbia Astrophysics Laboratory, Columbia University, New
 York, NY 10027}

\altaffilmark{8}{Laboratorio de Astrof\`{\i}sica Espacial y F\`{\i}sica
 Fundamental, P.O. Box 50727, 28080 Madrid, Spain}

\altaffilmark{9}{Istituto Fisica Cosmica e Applicazioni all'Informatica, 
C.N.R., Via U. La Malfa 153, 90146 Palermo, Italy}

\altaffilmark{10}{University of California, Berkeley, Space Science Laboratory, 
 Berkeley, CA 94720-7450, USA}

\altaffilmark{11}{Astrophysics Division, Space Science Department of ESA,
 ESTEC, Noordwijk, The Netherlands}

\begin{abstract}
We report high-energy spectral data of the prompt emission of GRB970228 and its
X--ray afterglow. We establish that the nature of the X--ray afterglow emission
is non-thermal and similar to the later portion of GRB970228. Our data can be
used to discriminate different emission models of GRB afterglows. While cooling
of excited compact objects can be ruled out, fireball models are constrained in
the physics of the radiation emission processes and their evolution.
\end{abstract}

\keywords{gamma rays: bursts --- gamma rays: observations ---  stars:
neutron --- X--rays: general}

\section{Introduction}

% Observations of celestial gamma--ray bursts (GRB) performed over the last 25
% years did not succeed in finding the origin of their emission. The detection
% and monitoring of their afterglows at other wavelengths can provide the
% information needed to determine the site and the physics of the GRB phenomenon.
Recent observations of the GRB970228 afterglow \cite{Costa97a} by the \S\
satellite \cite{Boella97} have shown that it may be actually possible to
identify the ultimate source and nature of emission of gamma--ray bursts
(GRBs). Eight hours after the burst, we detected a new transient X--ray source,
1SAX J0501.7+1146, by re-pointing the \S\ Narrow Field Instruments (NFIs) at
the GRB location. Various arguments render compelling the interpretation of
this source in terms of a fading X--ray afterglow from GRB970228: (a) the
backward extrapolation of the source light curve to the time of the burst is
consistent with the average X--ray flux detected during the late phase of the
burst \cite{Costa97b}; (b) its celestial position is consistent with the burst
position determined by the Interplanetary Network \cite{Hurley97}; (c) a fading
optical transient consistent with both the Wide Field Camera (WFC) and NFIs
positions was also discovered \cite{vanparadijs,guarnieri}; (d) an
observation of the GRB error box with {\em ROSAT\/} provided the position of
the X--ray transient source within a $10''$ radius. Its centroid
coincides, within $2''$, with the optical transient position
\cite{rosat,Frontera97c}. The optical transient appears to be embedded in an
apparently constant extended source \cite{Sahu97}. If it is a galaxy,
those models that place GRBs at cosmological distances would be  favoured.

In another paper \cite{Costa97b} we reported the decay of the light curve of
the source, which is described by a power law function ($\propto t^{-\alpha}$
with index $\alpha = 1.33^{+0.13}_{-0.11}$) that is consistent with that
expected in simple relativistically expanding fireball models \cite{Wijers97}.
%We also showed that a substantial fraction of the energy involved in the
%initial event is emitted as afterglow at later times (hours--days).
In this paper we report on the spectral evolution of GRB970228 and its
afterglow.

% This paper is structured as follows: in Section~2 we summarize the
% observational phases that led to the discovery of the GRB970228 X--ray
% afterglow and report on the main features of the initial event; in Section~3 we
% present the spectral properties of the burst and its X--ray afterglow; in
% Section~4 we discuss these results and draw some conclusions.

\section{Observations} 

GRB970228 was detected by the Gamma--Ray Burst Monitor (GRBM, 40--700 keV)
\cite{Frontera97} and WFC No. 1 (1.5--26.1~keV) \cite{Jager97}
aboard \S\ on February 28.123620 UT \cite{Costa97a}. Its position was
determined to be within an error radius of $3'$ (3$\sigma$) centered at
$\alpha_{2000} = 05^{\rm h}01^{\rm m}57^{\rm s}$, $\delta_{2000} =
11^\circ46'24''$. 
Eight hours after the initial detection, the Narrow Field
Instruments on-board \S\ were pointed  at the burst location for a first target
of opportunity (TOO1) observation from February 28.4681 to February
28.8330~UT. A new X--ray source, 1SAX J0501.7+1146, was detected \cite{Costa97b}
in the GRB error box by both the Low Energy (0.1--10 keV, 8725~s exposure time)
and Medium Energy (2--10~keV, 13444~s) Concentrators Spectrometers (LECS and
MECS). The same field was again observed about three
days later for a second observation (TOO2) from March 3.7345 to March
4.1174 (16270~s MECS and 9510~s LECS). During
this observation, the 2--10 keV source flux had decreased by about a factor 20.
The main arguments for the association of this source with the X--ray afterglow
of GRB970228 have been discussed in Section~1.

The data available from GRBM for spectral analysis  include two 1~s ratemeters
(40--700~keV and $>$100~keV) and 128~s  count spectra
(40--700~keV, 225 channels).
The energy resolution of the GRBM unit 1, co-aligned with the WFC No. 1, 
is 20\% at 280 keV \cite{Amati97}. 
WFCs (energy resolution $\approx$ 20\% at 6~keV) were operated in normal
mode with 31 channels in 1.5--26~keV and 0.5~ms time resolution \cite{Jager97}. 
%, in which each event is labelled with  position in the detector
%and energy ( ). The event time is given
%every each fourth detected photon with a resolution of 0.5~ms.   The energy
%resolution of the instrument is $\approx$ 20\% at 6~keV
The burst direction was offset by 14$^\circ$ with respect to
the WFC axis. The  effective area of the GRBM unit 1, co-aligned with the 
WFC No. 1, is $\approx$~420~cm$^2$ 
in the 40--700~keV band and is $\approx$~500~cm$^2$ at 300~keV.  The 
corresponding effective area of WFC No.~1 averaged in the 2--26~keV energy band
is 118~cm$^2$.
The background level in the WFC and GRBM energy bands  was variable during the
burst \cite{Feroci97}. 
This variation was apparent up to 100~keV, while above
100~keV the background was stable. 
For the WFC spectra and light curves, the
background level was estimated using an equivalent section of
the detector area not illuminated by the burst. In the case of GRBM, 
background during the burst was  estimated by interpolation, using a
quadratic function that fit the 150~s count rate data  before the burst 
and the 100~s data after its end. The presence of a variable
background  prevented the derivation of useful upper limits to the afterglow
emission after the burst.
During TOO1, the X-ray source associated with the burst afterglow was 
offset by about 10' with respect to the telescope axes. Since the image
centroid, for MECS 1 and 2, coincided with of a strong-back that
supports the detector window and absorbs most photons up to about 6~keV,  
we used for the spectral analysis only MECS 3 and LECS. The LECS
and MECS spectra for TOO1 were extracted from a $4'$ radius region
around the centroid of the image source. 
For TOO2, the source was on-axis and the
spectra from LECS and MECS were extracted from a $2'$ radius region. The
MECS spectra were equalized and
%the energy--channel relationship of the MECS 1 
added together. The background spectrum was estimated from long observations
(200~ks for MECS and 100~ks for LECS) of blank sky fields, using an equivalent
detector region of the source image with a similar offset. 
%The background level
%estimated in this way was in agreement with that obtained during TOOs, which
%however had lower statistics.

Figure~\ref{fig:timeprofile} shows the time profiles of GRB970228 in different
energy bands after the background subtraction. In the $\gamma$--ray energy band
(40--700~keV), the burst is characterized by an initial 5~s strong pulse
followed, after 30~s, by a set of three additional pulses of decreasing
intensity. In the X--ray band (1.5--10~keV) the first pulse starts within one
second of the $\gamma$--ray pulse, achieves the peak flux later and has a
duration three times longer. The entire burst duration is $\sim 80$~s. The
$\gamma$--ray (40--700~keV) fluence of the entire burst is $(1.1\pm 0.1)\times
10^{-5} \ergcm$, which is a factor $5\pm 1$ higher than the corresponding value
in the 2--10 keV band. (This ratio is based on the more accurate analysis of
the WFC data performed for this paper and corrects the approximate value given
previously by Costa et~al.\ 1997c) \nocite{Costa97c}. For comparison, the
40--700~keV fluence integrated over the first pulse is $(6.1\pm 0.2)\times
10^{-6} \ergcm$, which is a factor of $8.5\pm 1.3$ higher than the
corresponding fluence in 2--10~keV. The $\gamma$--ray peak flux is $(3.7\pm
0.1)\times 10^{-6} \ergcms$, while the corresponding 2--10 keV flux is $(1.4\pm
0.1)\times 10^{-7} \ergcms$, with a ratio between $\gamma$--ray and X--ray peak
fluxes of $26\pm 2$. 
In the BATSE energy band (50--300~keV) the fluence of
GRB970228 is 6.1$\times 10^{-6} \ergcm$.

\section{Spectral analysis and results}

We divided the GRB time profile into seven temporal sections, and performed  a
separate spectral analysis on the average spectrum of each section. The
duration of the sections were chosen on the basis of the statistical quality of
the data. In particular, we obtained two spectra (5 and 9~s duration,
respectively) for the 1$^{st}$ pulse (A and B in Fig.~\ref{fig:timeprofile}), 
four 3~s spectra for the $2^{nd}$ pulse (C, D, E and F), and only one spectrum
for the $3^{rd}$ and $4^{th}$ pulses together 
(G in Fig.~\ref{fig:timeprofile}). All
WFC spectra are well fit with power laws ($K_X\cdot E^{\alpha_X}$), while
black-body spectra do not fit the data. We assumed power laws ($K_\Gamma\cdot
E^{\alpha_\Gamma}$) to deconvolve spectral data in the GRBM band. In
Fig.~\ref{fig:power} we show the logarithmic power per photon energy decade
(the $\nu F_\nu$ spectrum) for six of the time intervals defined above.
The spectral evolution of the 1$^{st}$  GRB pulse and its diversity from the
subsequent pulses is evident. The initial part of the 1$^{st}$ pulse
(Fig.~\ref{fig:power}A) is quite hard, with noticeable curvature of the $\nu
F_\nu$ spectrum. The peak energy, that is not detectable in the WFC/GRBM 
energy range, is likely above 700~keV.  The decay part of the 1$^{st}$ pulse 
shows
substantial evolution, with a clearly detectable broad curvature. By fitting
the photon spectrum with a smoothly broken power law \cite{Band93}, we found a
peak energy at E$_p = 35\pm 18$~keV. This peak energy behaviour  is typical of
GRB spectral evolution \cite{ford95}.
In Fig.~\ref{fig:evolution} we show the time evolution from the burst onset of
the GRBM and WFC power law spectral indices $\alpha_\Gamma$ and $\alpha_X$. For
the 1$^{st}$ pulse, the temporal behaviour of $\alpha_\Gamma$  is well described
by a linear law ($\alpha_\Gamma(t) = a+bt$, $\chi^2 = 1.28$, 3 dof) with
$a=-0.9\pm 0.2$ and $b=-0.31\pm 0.05$, and t in seconds. The power law photon
indices $\alpha_\Gamma$ and $\alpha_X$ do not evolve in the same way as a
consequence of the curvature of the 1$^{st}$ pulse spectra. We also note the
interesting spectral softening during the decaying part of the pulse.

The spectral behaviour of the $2^{nd}$ pulse is significantly different 
from that
of the 1$^{st}$ one. The  spectral data can be fit with a single power law over
the 1.5--700~keV range. During the first 3~s of the $2^{nd}$ pulse the
spectrum is significantly harder than during the last part of the 1$^{st}$ 
pulse.
The photon index $\alpha_\Gamma$ of the $2^{nd}$ pulse vs.\ time from 
pulse onset
shows some evidence of evolution (it can be fit with a linear law as above with
$a=-1.6\pm 0.2$ and $b =-0.12\pm0.04$), but it is also consistent with a
constant value. Instead $\alpha_X$ is definitely constant during the $2^{nd}$ 
pulse. The overall WFC/GRBM photon index of the $2^{nd}$ pulse is initially
$-1.8\pm 0.1$ and then $-1.9\pm 0.1$. Interestingly, there is no strong
evidence for spectral curvature as shown in Fig.~\ref{fig:power}, probably
indicating that the peak energy $E_p$ passed through the WFC/GRBM energy band
during the first part of the burst and, as a consequence, later it is  below
$\sim 2$~keV.
Low counting statistics prevents us from deriving good quality spectra for the
last section of the $2^{nd}$ pulse (section F) and for the $3^{rd}$ and 
$4^{th}$ 
pulses, separately. For section F (not shown in Fig.~\ref{fig:power}), the
X--ray data were fit by a power-law model with $\alpha_X = -1.5\pm 0.4$, while
the $\gamma$--ray data provided only a 2$\sigma$ upper limit of the
$\alpha_\Gamma$ spectral index of $-0.6$. The average photon spectrum of the
sum of the $3^{rd}$ and $4^{th}$ is well fit with a power-law model of 
$\alpha_\Gamma = -1.4^{+0.5}_{-0.3}$ and $\alpha_X = -1.6\pm 0.1$. It appears
that the spectrum is again harder than that at the end of the $2^{nd}$ pulse and
more similar to the one at the beginning of the $2^{nd}$ pulse. In any case, it
is consistent with a single power-law. The $\nu F_\nu$ spectrum of the sum of
the $3^{rd}$ plus $4^{th}$ pulses is shown in Fig.~\ref{fig:power}G.

Remarkably, we find that the power-law spectral model continues to hold also
during the X--ray afterglow. The photon spectrum of 1SAX J0501.7+1146 in the
0.1--10~keV energy band as measured during the TOO1 observation is shown in
Fig.~\ref{fig:spectrum}. It was obtained from the source count spectra of LECS
and MECS after deconvolution for the
instrument responses. Spectral fits were performed with the XSPEC 9.0 package. 
We fit various functions to the source count rate spectrum (see
Table~\ref{table}). The best fit model, with a $\chi^2 = 7.2$ (10 dof) is a
power-law of photon index $\alpha = 2.06\pm 0.24$ and photoelectric absorption
${\rm N}_H = 3.5^{+3.3}_{-2.3}\times 10^{21}$~cm$^{-2}$. A
power-law model with ${\rm N}_H = 0$ does not fit the data nor does an absorbed
black-body model. The probability that the difference in the $\chi^2$ between
these models and the absorbed power-law is due to chance is $1.2\times 10^{-3}$
and $7.2\times 10^{-4}$, for the power-law with ${\rm N}_H = 0$ and for the
absorbed black-body, respectively. Our determination of ${\rm N}_H$ is
consistent with the total Galactic photoelectric absorption
($\sim$~$1.6\times 10^{21}$~cm$^{-2}$) expected along the line of
sight to the GRB error box. We can exclude a nearby source with
zero equivalent ${\rm N}_H$ (\eg near the Solar system) at about $3\sigma$
statistical significance. In Fig.~\ref{fig:spectrum}, we also show the best
fit model spectrum of the X--ray afterglow.
During the TOO2 observation, the X--ray source was too weak to derive a
high-quality spectrum. The MECS detected a 2--10~keV flux of $(1.5\pm
0.5)\times 10^{-12} \ergcms$ from the source, while the LECS did not detect any
significant flux ($3\sigma$ upper limit of $0.4\times 10^{-12} \ergcms$). The
hardness ratio HR$_1$ between the count rate in the hard X--ray band
(2--10~keV) and that in the soft band (0.1--2~keV) is $2.5\pm 0.8$ for TOO1 and
$>$2.5 for TOO2 ($3\sigma$ lower limit). However the hardness ratio HR$_2$
between two hard X--ray bands (3--5~keV and 1.5--3~keV) does not show
statistically significant variations from the first to the second TOO
observation. Thus we cannot infer a change in the spectral shape from TOO1 to
TOO2.

\section{Discussion and conclusions}
The non-thermal spectrum of the burst radiation and of the X--ray afterglow is
the most important result of this analysis. A power law model was found to fit
the spectral data better than a black body model. The maximum value of
$\alpha_\Gamma$ ($-0.9\pm 0.2$) achieved at the onset of the 1$^{st}$ pulse as
well as the corresponding $\alpha_X$ during the first 5~s are consistent with
the asymptotic spectral index below the peak energy E$_p$ ($-0.67$), as
expected in synchrotron emission models (\eg the Synchrotron Shock Model (SSM)
\cite{Tavani96}). The spectral index of the 1$^{st}$ pulse rapidly evolves to
$\alpha_\Gamma = -2.3\pm 0.3$ at the end of the $\gamma$--ray pulse. In the SSM
framework the final value of $\alpha_\Gamma$ during the 1$^{st}$ pulse 
corresponds
to an index of the non-thermal electron energy distribution function $\delta =
3.7$.
During the first 3~s of the $2^{nd}$ pulse the spectrum is significantly harder
than during the last part of the 1$^{st}$ pulse. This indicates significant
re-energization or relaxation of the particle energy distribution function
within the $\sim 20$~s between the 1$^{st}$ and the $2^{nd}$ pulse.
It appears that the last three GRB pulses and the X--ray afterglow have a
similar non-thermal spectrum and that this spectrum does not appear to change
from the first to the second TOO. These results are not consistent with simple
cooling models of excited compact objects. An analysis of {\em Ginga\/} data
suggested the existence of black-body spectral components in the `precursor' or
`delayed' GRB emission in the 1--10~keV band \cite{Yoshida89,Murakami91}. In
this burst we find no evidence for a black-body spectral component of
temperature $\sim 1$--2~keV or for a prominent soft X--ray component.
Furthermore, there is no evidence of upturn in the soft X--ray intensity with
respect to the higher energy spectrum \cite{Liang97}.

An evolving non-thermal spectrum for both the burst and afterglow emission is
generally expected in relativistic expanding fireball models
\cite{Meszaros97,Tavani97}. In these models, the physics and locations of the
shocks associated with the fireballs heavily influence the photon emission
mechanisms. Simple fireball models, in which only the forward blast wave
radiates efficiently, predict an evolution of the peak energy as a function of
the time $t$ from the burst onset, $E_p \propto t^{-3/2}$ 
\cite{Tavani97,Wijers97}. By extrapolating our data for the 1$^{st}$ pulse of
GRB970228 we obtain an initial $E_p \simeq 1$~MeV. In this model, the overall
evolution of the X--ray intensity is expected to evolve as $\propto t^\delta$
with $\delta = (3/2)(\alpha + 1)$ and $\alpha$ an appropriate photon index
\cite{Wijers97}. If we use our best value of $\alpha_\Gamma$ for the decay part
of the $2^{nd}$ pulse ($-1.94\pm 0.13$), we obtain $\delta = (-1.4\pm 0.2)$, a
value that is  consistent with the observed decay of the GRB970228 afterglow
\cite{Costa97b}. However, if we use the value of $\alpha_\Gamma$ determined at
the end of the 1$^{st}$ pulse ($-2.3\pm 0.3$), the resulting value of $\delta$
($-1.95\pm 0.45$) would not agree with our observations. The discontinuity in
the $\gamma$--ray spectral index observed from the end of the 1$^{st}$ pulse to
the beginning of the $2^{nd}$ pulse requires an interpretation. Given the
continuity of the spectral index, starting from the $2^{nd}$ pulse, any physical
relation between the X--ray component of the GRB and the afterglow emission
most likely holds with the last set of hard GRB pulses, not with the 1$^{st}$ 
one.
This suggests that the emission mechanism producing the X--ray afterglow might
be already taking place after the 1$^{st}$ pulse.

\acknowledgements

This research is supported by the Italian Space Agency (ASI). 
%We thank the \S\
%mission director R.C.\ Butler and both teams of the \S\ Operative Control
%Center and Scientific Data Center.
% for their efficient and enthusiastic support
% to the GRB alert program. 
KH is grateful to the US 1SAX Guest Investigator program for support.

\newpage

\begin{deluxetable}{lcccccl}
\tablewidth{0pt}
\tablecaption{Spectral fits of 1SAX J0501.7+1146 in the energy range 
0.1--10~keV \tablenotemark{a} \label{table}}
\tablehead{
\colhead{Model} &
\colhead{A \tablenotemark{b}} &
\colhead{$\alpha$} &
\colhead{kT (keV)} &
\colhead{N$_H (\times 10^{21}$ cm$^{-2})$} &
\colhead{$\chi^2$} &
\colhead{dof}}
\startdata
Power Law plus abs\ \tablenotemark{c} & $1.19^{+0.87}_{-0.50}$ & $2.06\pm0.24$ & \nodata     & $3.5^{+3.3}_{-2.3}$ & 7.2  & 10 \nl
PL with abs set to 0                & $0.52\pm0.15$          & $0.45\pm0.20$ & \nodata     & 0 (frozen)          & 17.8 & 11 \nl
Black Body plus abs\ \tablenotemark{c}& $0.032\pm0.005$        & \nodata       & $0.86\pm0.12$ & $0^{+1}$          & 18.7 & 10 \nl
\tablenotetext{a}{\ Uncertainties at the 90\% confidence level for a single
parameter.}
\tablenotetext{b}{\ Photons cm$^{-2}$ s$^{-1}$ keV$^{-1}$ at 1 keV.}
\tablenotetext{c}{\ Photoelectric absorption of a gas with cosmic abundance
\cite{Morrison}.}
\enddata
\end{deluxetable}

\ifPreprintMode\relax\else

\newpage

% Figure 1
\figcaption[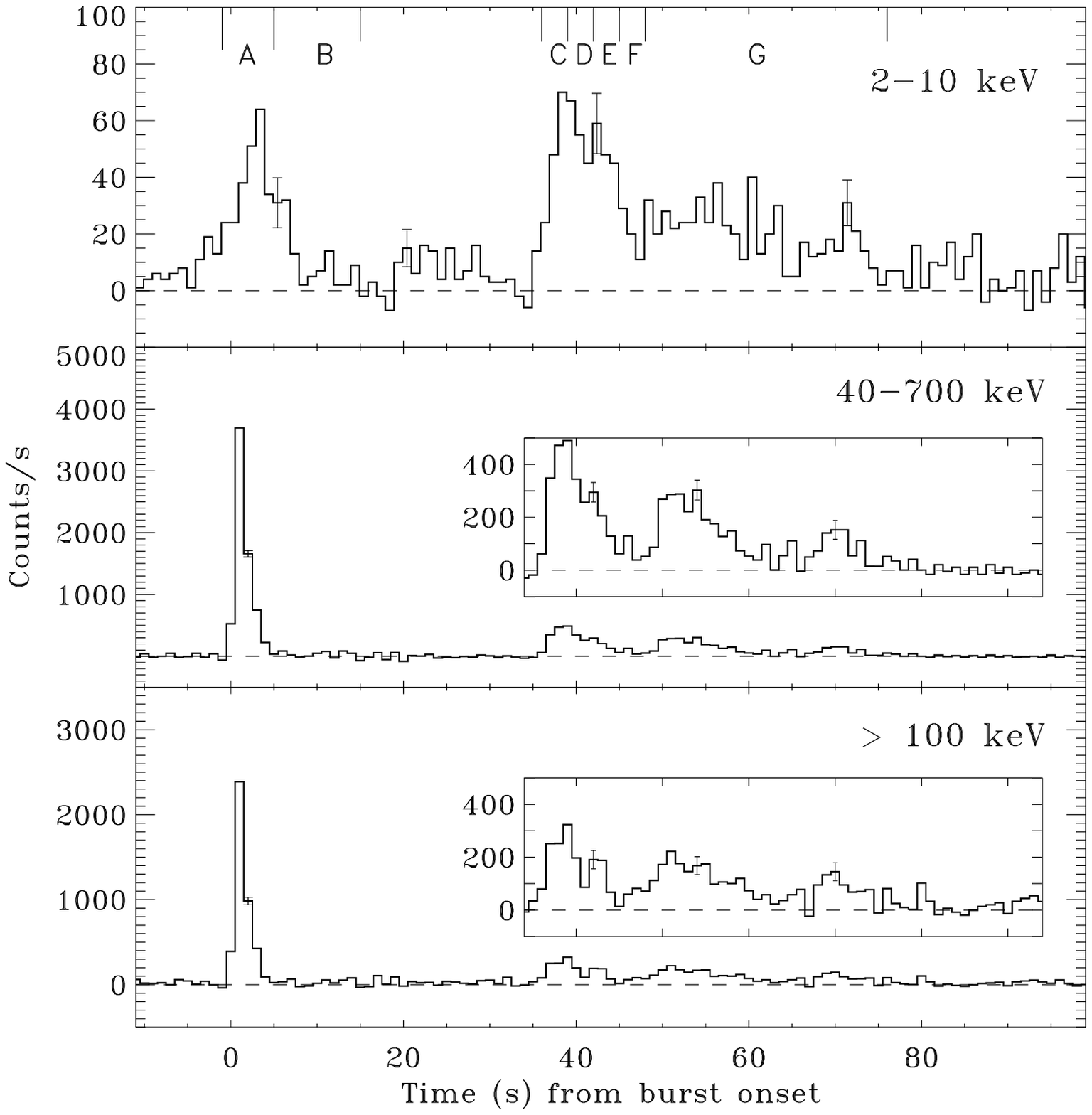]{Light curve of GRB970228 in different energy bands,
after background subtraction (see text). Dashed lines give the background
reference. In the insets, the three minor pulses of the burst. The time
sections on which the spectral analysis was performed are shown on the
top. \label{fig:timeprofile}}

% Figure 2
\figcaption[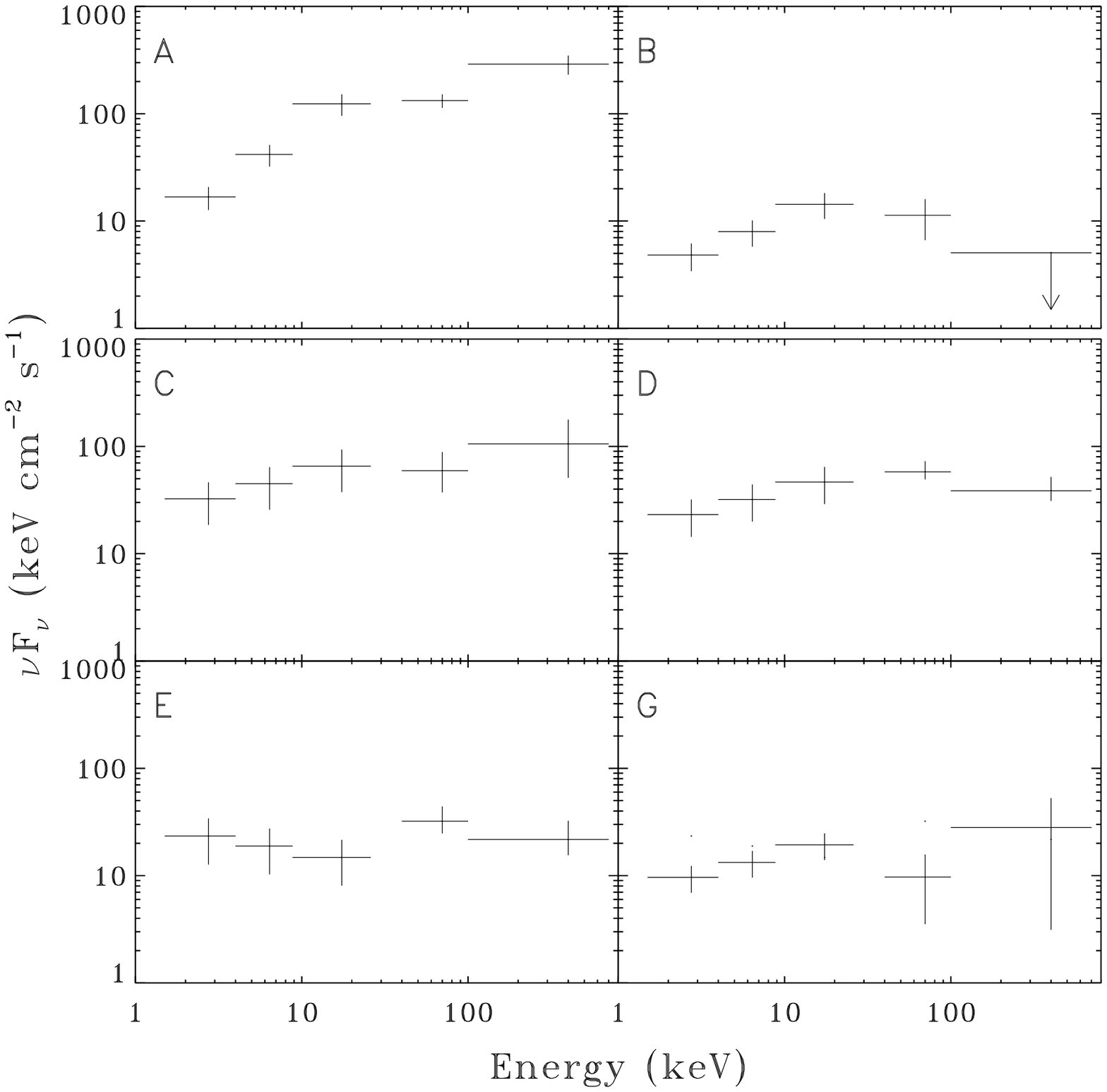]{$\nu F_\nu$ spectrum of the burst ($\nu$ is the photon
energy in keV and $F_\nu$ is the specific energy flux in
keV~cm$^{-2}$~s$^{-1}$~keV$^{-1}$) for 6 of the 7 time sections in which we
divided the burst time profile. The time intervals from the burst onset over
which the spectrum has been accumulated are also shown. \label{fig:power}}

% Figure 3
\figcaption[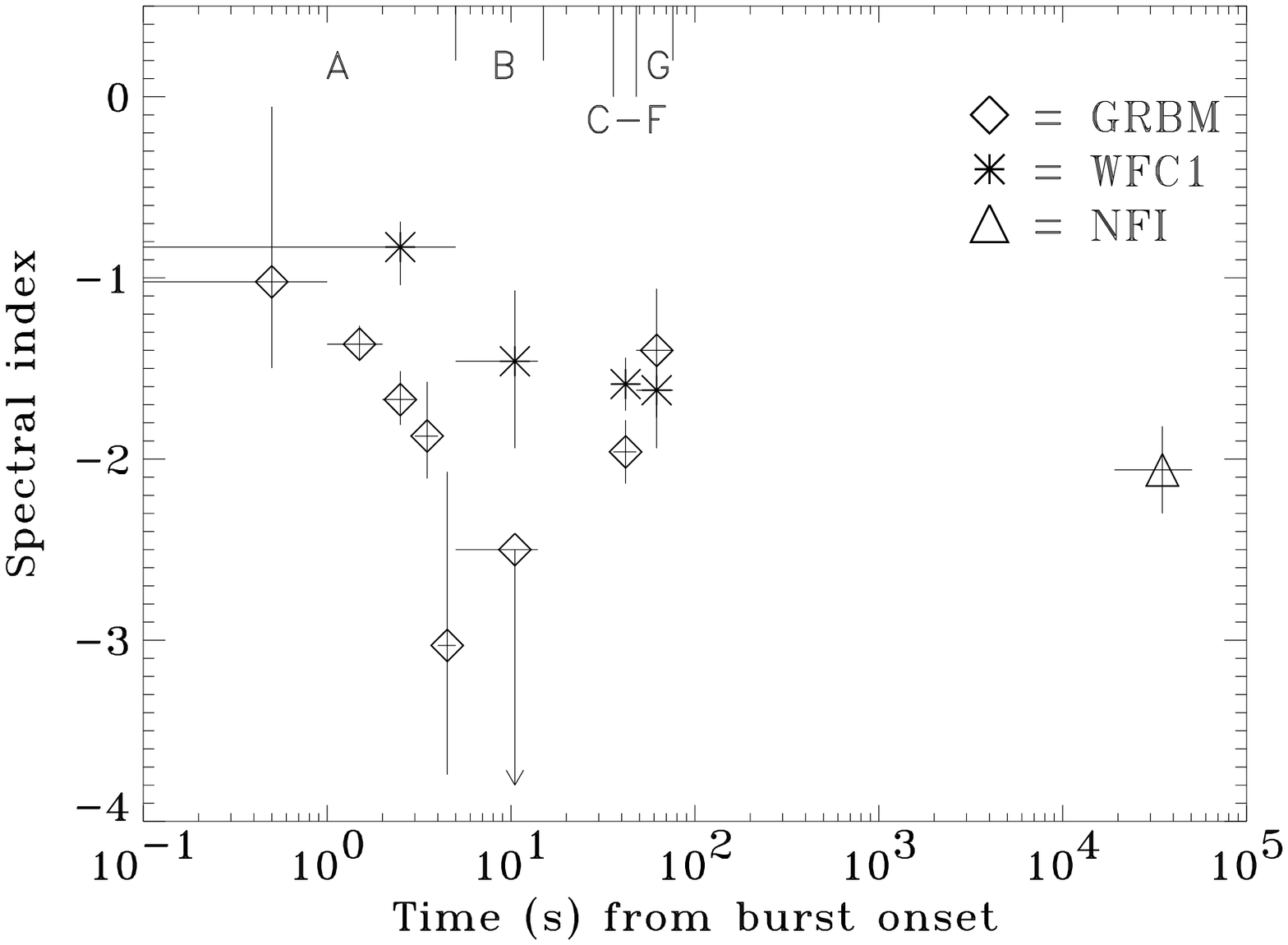]{Behavior of the X--ray and $\gamma$--ray power law photon
indices with time from the burst onset. The photon index of the X--ray
afterglow spectrum is also shown. \label{fig:evolution}}

% Figure 4
\figcaption[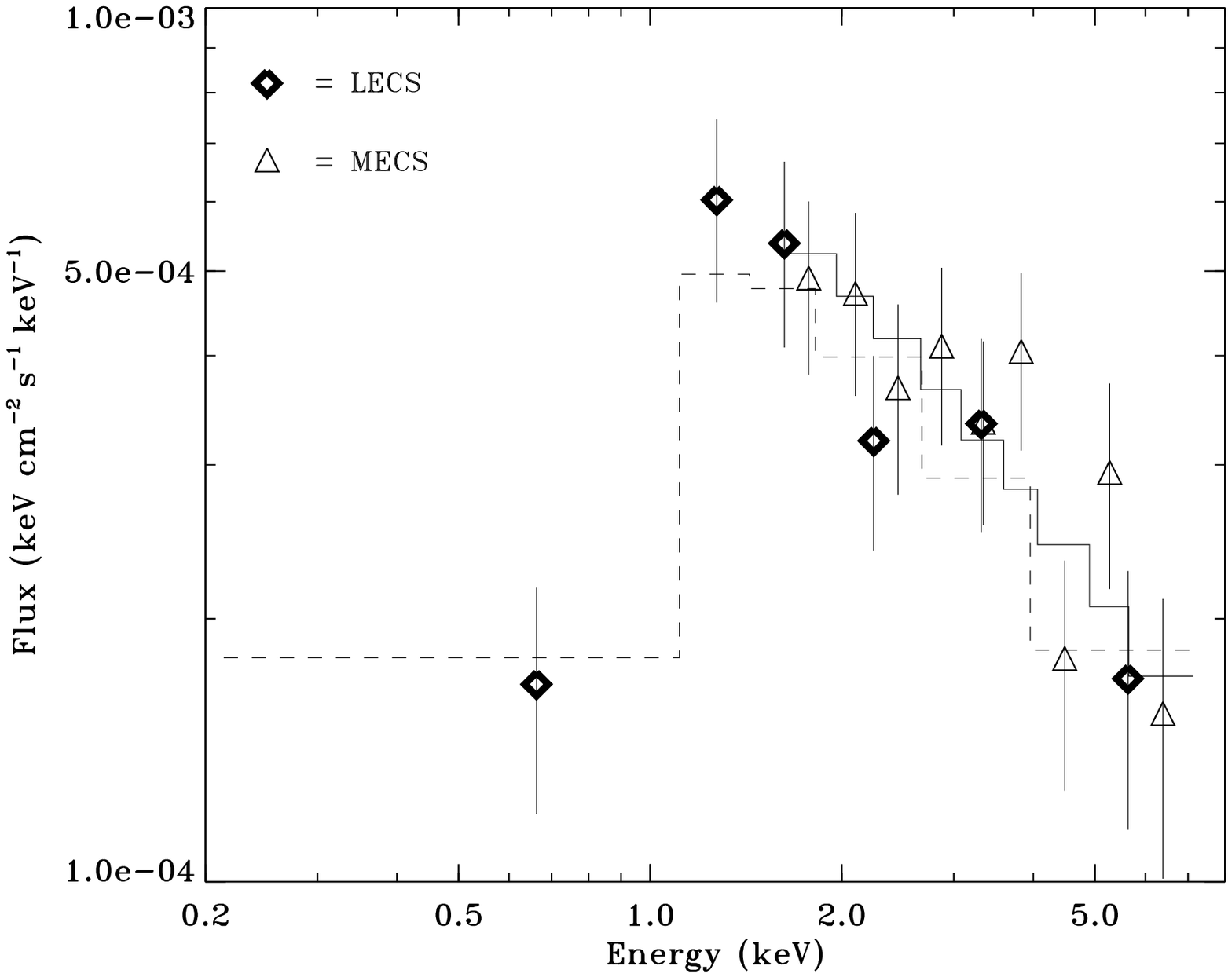]{Deconvolved LECS and MECS spectra of the X--ray afterglow
during TOO1. The best fit power law model with low energy photoelectric
absorption is superposed to the LECS (dashed line) and MECS (full line) data.
\label{fig:spectrum}}

\fi

\newpage

\begin{figure}
\plotone{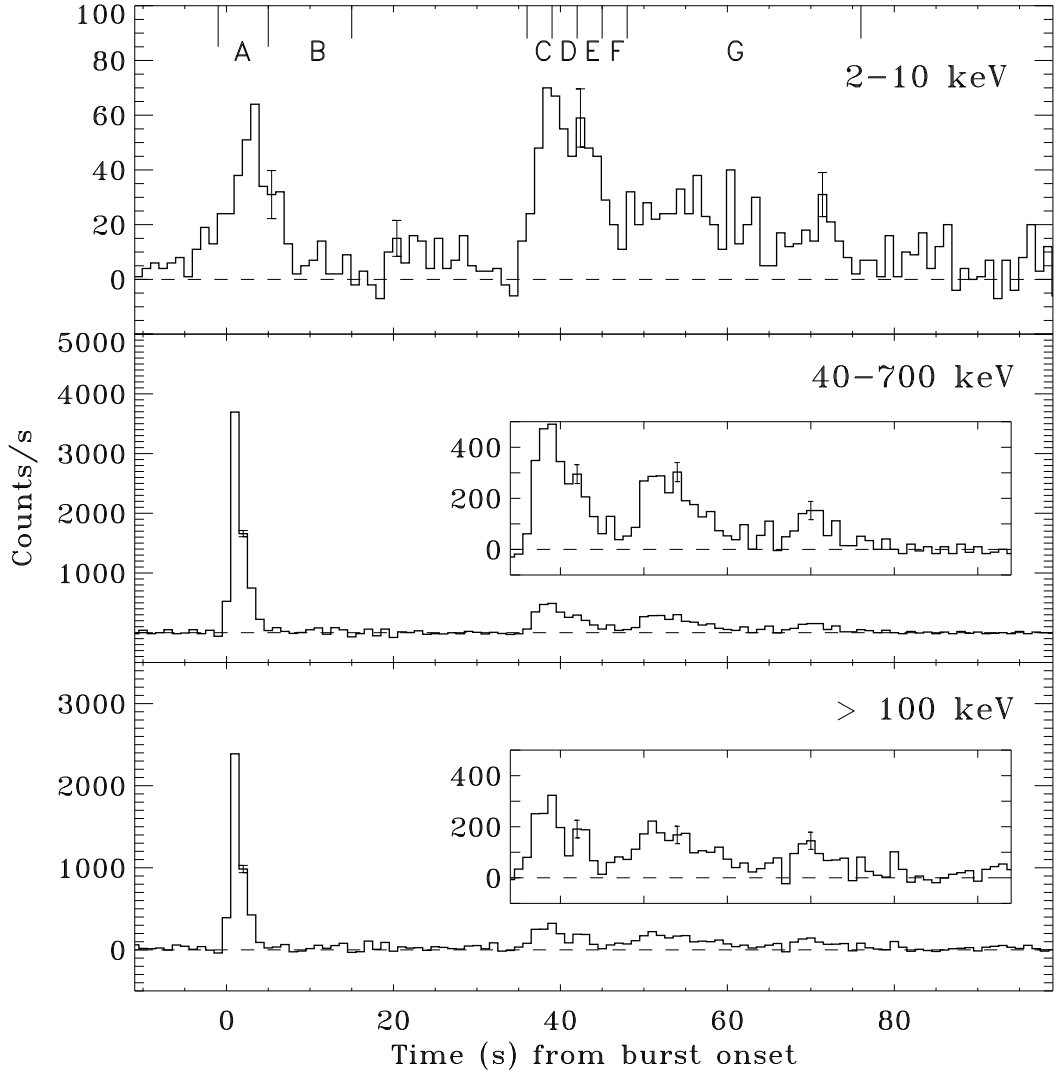}
\ifPreprintMode
\vspace{2cm}
\caption[]{Light curve of GRB970228 in different energy bands, after background
subtraction (see text). Dashed lines give the background reference. In the
insets, the three minor pulses of the burst. The time sections on which the
spectral analysis was performed are shown on the top.}
\label{fig:timeprofile}
\fi
\end{figure}

\begin{figure}
\plotone{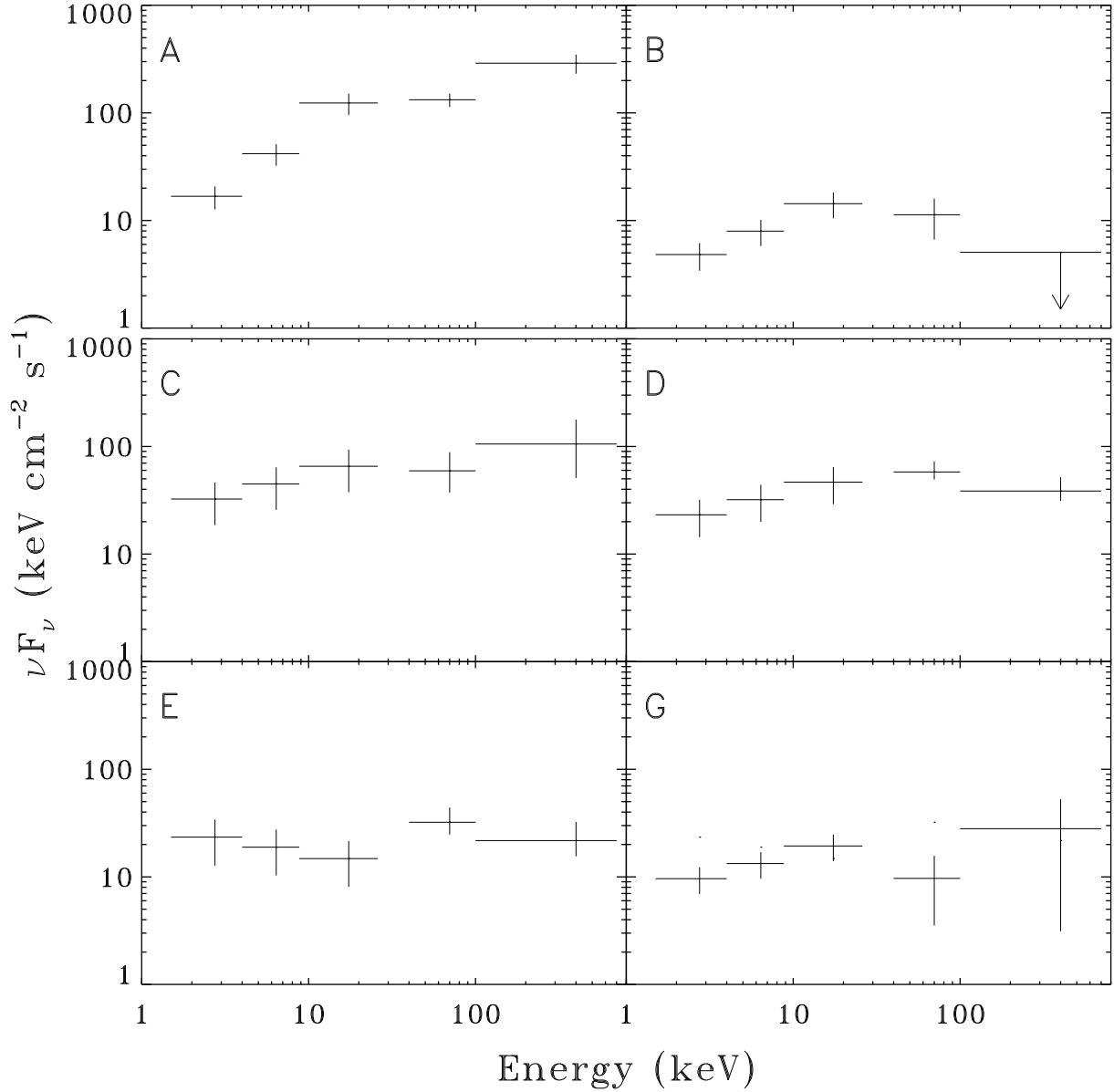}
\ifPreprintMode
\vspace{2cm}
\caption[]{$\nu F_\nu$ spectrum of the burst ($\nu$ is the photon
energy in keV and $F_\nu$ is the specific energy flux in
keV~cm$^{-2}$~s$^{-1}$~keV$^{-1}$) for 6 of the 7 time sections in which we
divided the burst time profile. The time intervals from the burst onset over
which the spectrum has been accumulated are also shown.}
\label{fig:power}
\fi
\end{figure}

\begin{figure}
\plotone{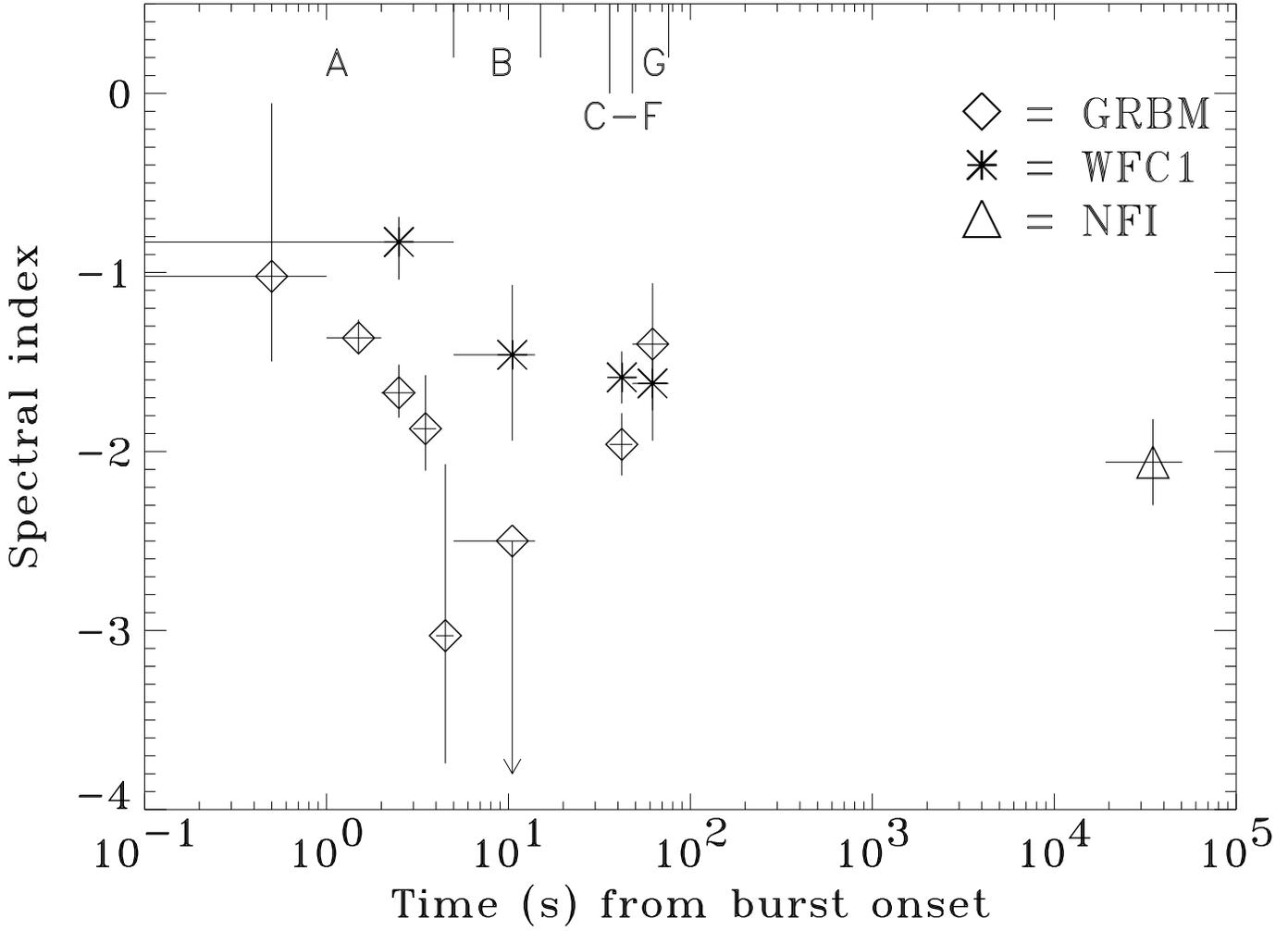}
\ifPreprintMode
\caption[]{Behavior of the X--ray and $\gamma$--ray power law photon
indices with time from the burst onset. The photon index of the X--ray
afterglow spectrum is also shown.}
\label{fig:evolution}
\fi
\end{figure}

\begin{figure}
\plotone{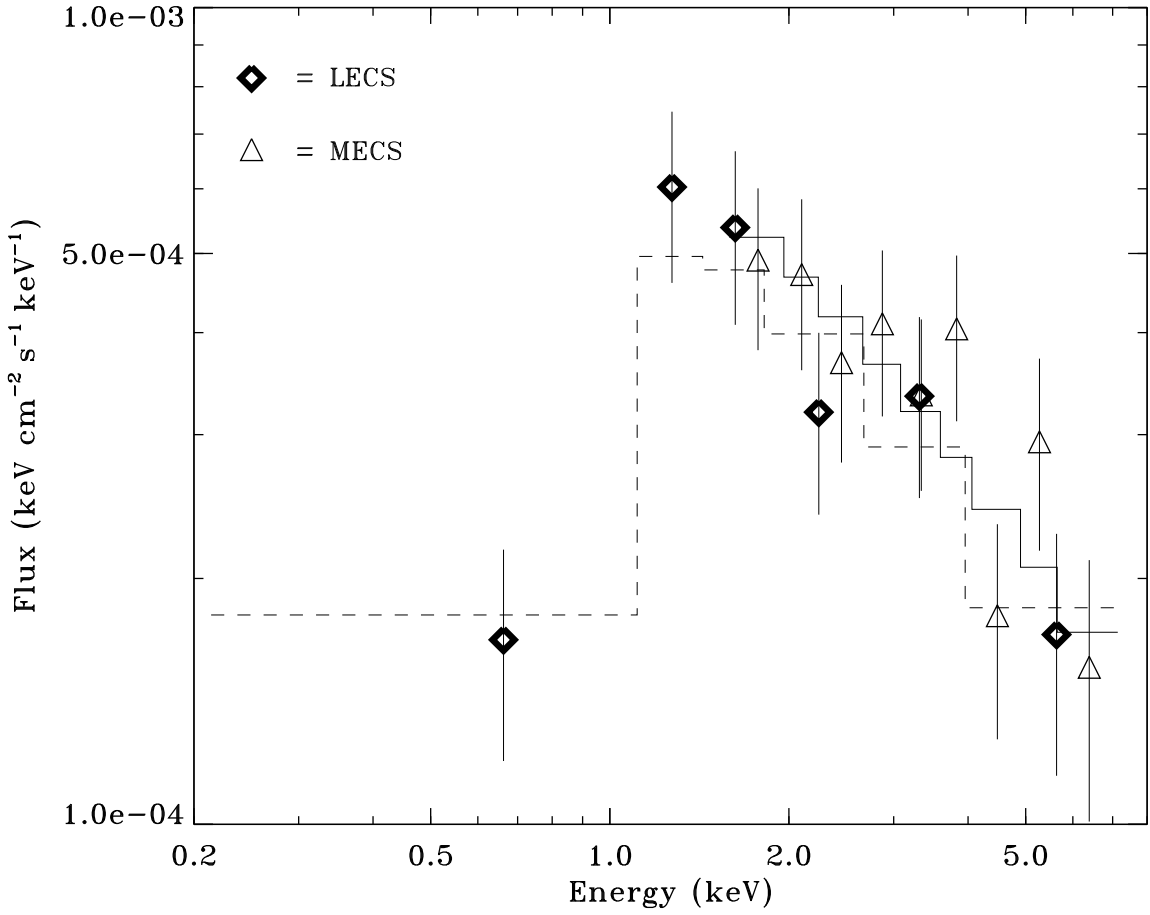}
\ifPreprintMode
\caption[]{Deconvolved LECS and MECS spectra of the X--ray afterglow
during TOO1. The best fit power law model with low energy photoelectric
absorption is superposed to the LECS (dashed line) and MECS (full line) data.}
\label{fig:spectrum}
\fi
\end{figure}


\begin{thebibliography}{99}

\bibitem[Amati et~al.\ 1997]{Amati97}
Amati, L. et~al. 1997, SPIE Proceedings, 3114, in press

%\bibitem[Arnaud 1996]{Arnaud}
%Arnaud, K.A. 1996, in Proc. V ADASS Symposium, ASP Conf. Series,
%Vol. 101, eds. Jacoby, G.H. and Barnes, J., (Astronomical Society of Pacific),
%p.17

\bibitem[Band et~al.\ 1993]{Band93}
Band, D. et~al. 1993, \apj, 413, 281

\bibitem[Boella et~al.\ 1997]{Boella97}
Boella, G. et~al. 1997a, \aaps,  122, 299

%\bibitem[Boella et~al.\ 1997b]{Boella97b}
%Boella, G. et~al. 1997b, \aaps, 122, 327
%
\bibitem[Costa et~al.\ 1997a]{Costa97a}
Costa, E. et~al. 1997a, \iaucirc, 6572 
%
%\bibitem[Costa et~al.\ 1997b]{Costa97b}
%Costa, E. et~al. 1997b, \iaucirc, 6576

\bibitem[Costa et~al.\ 1997b]{Costa97b}
Costa, E. et~al. 1997b, \nat, 387, 783

%\bibitem[Dickey and Lockman 1990]{Dickey90}
%Dickey, J.M. and Lockman, F.J. 1990, \araa, 28, 215

\bibitem[Feroci et~al.\ 1997]{Feroci97}
Feroci, M. et~al. 1997, SPIE Proceedings, 3114, in press

\bibitem[Ford et~al.\ 1995]{ford95}
Ford, L.A., et~al. 1995, \apj, 439, 307

\bibitem[Frontera et~al.\ 1997a]{Frontera97}
Frontera, F., Costa, E., Dal~Fiume, D., Feroci, M., Nicastro, L., Orlandini,
  M., Palazzi, E., and Zavattini, G. 1997a, \aaps, 122, 357

\bibitem[Frontera et~al.\ 1997b]{rosat}
Frontera, F. et~al. 1997b, \iaucirc, 6637

\bibitem[Frontera et~al.\ 1997c]{Frontera97c}
Frontera, F. et~al. 1997c, in preparation.

%\bibitem[Galama et~al.\ 1997]{Galama97}
%Galama, T. et~al. 1997, \nat, 387, 479

\bibitem[Guarnieri et~al.\ 1997]{guarnieri}
Guarnieri, A. et~al. 1997, \aap, in press

\bibitem[Hurley et~al.\ 1997]{Hurley97}
Hurley, K., et~al. 1997, \apjl, 485, L1 
\bibitem[Jager et~al.\ 1997]{Jager97}
Jager, R., et~al. 1997, \aaps, in  press

\bibitem[Liang et~al.\ 1997]{Liang97}
Liang, E., Kusunose, M., Smith, I.A. and Crider, A. 1997, \apjl, 479, L35

\bibitem[M\'esz\'aros and Rees 1997]{Meszaros97}
M\'esz\'aros, P. and Rees, M.J. 1997, \apj, 476, 232

\bibitem[Morrison and McCammon 1983]{Morrison}
Morrison R. and McCammon, D. 1983, \apj, 270, 119

\bibitem[Murakami et~al.\ 1991]{Murakami91}
Murakami, T., Inoue, H., Nishimura, J., van~Paradijs, J., Fenimore, E.E.,
Ulmer, A. and Yoshida, A. 1991, \nat, 350, 592

%\bibitem[Pamini et~al.\ 1990]{Pamini90}
%Pamini, M., Natalucci, L., Dal~Fiume, D., Frontera, F., Costa, E. and Salvati,
%M. 1990, Il Nuovo Cimento, 13C, 337
%
%\bibitem[Parmar et~al.\ 1997]{Parmar97}
%Parmar, A.N. et~al. 1997, \aaps, 122, 309

%\bibitem[Pedichini et~al.\ 1997]{pedichini}
%Pedichini, F. et~al. 1997, \iaucirc, 6635

\bibitem[Sahu et~al.\ 1997]{Sahu97}
Sahu, K.C. et~al. 1997, \nat, 387, 476

\bibitem[Tavani 1996]{Tavani96}
Tavani, M. 1996, \apj, 466, 768

\bibitem[Tavani 1997]{Tavani97}
Tavani, M. 1997, \apj, 483, L87

\bibitem[van~Paradijs et~al.\ 1997]{vanparadijs}
van~Paradijs, J. et~al. 1997,  \nat,  386, 686

\bibitem[Yoshida et~al.\ 1989]{Yoshida89}
Yoshida, A. et~al. 1989, \pasj, 41, 509

\bibitem[Wijers et~al.\ 1997]{Wijers97}
Wijers, R.A.M.J., Rees, M.J., and M\'esz\'aros, P. 1997, \mnras, in press.

\end{thebibliography}
\end{document}